\documentclass[prd,twocolumn,nofootinbib,superscriptaddress,amssymb,amsfonts,amsmath,preprintnumbers]{revtex4-2}

\usepackage{amsmath}

\usepackage[colorlinks=true,linktoc=page,linkcolor=purple,citecolor=teal,urlcolor=magenta]{hyperref}
\usepackage{url}

\usepackage{graphicx,tabularx,tabularray}

\usepackage[compat=1.1.0]{tikz-feynman}

\usepackage{xcolor}
\definecolor{dgreen}{rgb}{.0,.6,.0}
\definecolor{lime}{HTML}{A6CE39}
\definecolor{lg}{RGB}{220,220,220} 

\usepackage{natbib}

\usepackage{cancel}

\DeclareRobustCommand{\orcidicon}{\hspace{-2.1mm}
\begin{tikzpicture}
\draw[lime,fill=lime] (0,0.0) circle [radius=0.13] node[white] {{\fontfamily{qag}\selectfont \tiny ID}}; \draw[white,fill=white] (-0.0525,0.095) circle [radius=0.007]; 
\end{tikzpicture} \hspace{-3.7mm} }
\foreach \x in {A, ..., Z} {\expandafter\xdef\csname orcid\x\endcsname{\noexpand\href{https://orcid.org/\csname orcidauthor\x\endcsname} {\noexpand\orcidicon}}}

\newcommand{\Eprint}[1]{\href{#1}}

\begin{document}

\preprint{PSI-PR-24-01, ZU-TH 01/24, ICPP-78}

\title{Combined Explanation of LHC Multi-Lepton, Di-Photon and Top-Quark Excesses}


\author{Guglielmo Coloretti\orcidA{}}
\email{guglielmo.coloretti@physik.uzh.ch}
\affiliation{Physik-Institut, Universität Zürich, Winterthurerstrasse 190, CH–8057 Zürich, Switzerland}
\affiliation{Paul Scherrer Institut, CH–5232 Villigen PSI, Switzerland}

\author{Andreas Crivellin\orcidB{}}
\email{andreas.crivellin@cern.ch}
\affiliation{Physik-Institut, Universität Zürich, Winterthurerstrasse 190, CH–8057 Zürich, Switzerland}
\affiliation{Paul Scherrer Institut, CH–5232 Villigen PSI, Switzerland}

\author{Bruce Mellado}
\email{bmellado@mail.cern.ch}
\affiliation{School of Physics and Institute for Collider Particle Physics, University of the Witwatersrand,
Johannesburg, Wits 2050, South Africa}
\affiliation{iThemba LABS, National Research Foundation, PO Box 722, Somerset West 7129, South Africa}

\begin{abstract}

In this article, we propose the $\Delta$2HDMS as a combined explanation of several excesses and anomalies observed at the LHC. Especially, $t\bar t$ differential distributions point towards the associated production of new electroweak scale Higgs bosons decaying into bottom quarks and $W$ bosons ($>\!5\sigma$) with masses consistent with the di-photon excesses at $\approx\!\!95\,$GeV and $\approx\!\!151.5\,$GeV ($3.8\sigma$ and $4.9\sigma$, respectively). Furthermore, CMS found indications for resonant $t\bar{t}$ production at $\approx\!400$\,GeV ($3.5\sigma$) and both ATLAS and CMS reported elevated four-top and $t\bar t W$ cross-sections. 

The $\Delta$2HDMS is obtained by supplementing the SM Higgs ($H_2$) with a second scalar doublet ($H_1$), real scalar singlet ($S$), and a Higgs triplet with zero hypercharge ($\Delta$). We fix the masses of the neutral triplet-like and the singlet-like scalars by the di-photon excesses, i.e.~$m_{\Delta^0}=151.5\,$GeV and $m_{S}=95\,$GeV, respectively. $H$, the CP-even component of $H_1$, is produced via gluon fusion from a top-loop and decays dominantly to $S+\Delta^0$ whose subsequent decays to $W^+W^-$ and $b\bar b$ explain the differential $t\bar t$ distributions for $\sigma(pp\to H\to S\Delta^0)\approx\!5$pb. Choosing the top-Yukawa of $H_1$ accordingly, the CP-odd Higgs boson $A$ turns out to have the right production cross-section to account for the resonant $t\bar{t}$ excess at 400\,GeV, while the top-associated production of $H$ and $A$ results in new physics pollution of Standard Model $t\bar{t}W$ and $t\bar{t}t\bar{t}$ cross sections, as preferred by the data. 
\end{abstract}
\maketitle

\section{Introduction} 
The Standard Model (SM) of particle physics describes the known fundamental constituents of matter and their interactions (excluding gravity) at sub-atomic scales. It has been extensively tested and verified by a plethora of measurements~\cite{ParticleDataGroup:2020ssz}. However, the minimality of the SM Higgs sector~\cite{Higgs:1964ia,Englert:1964et,Higgs:1964pj,Guralnik:1964eu} remains unexplained: While we know that the observed 125$\,$GeV boson~\cite{Aad:2012tfa,Chatrchyan:2012ufa} has properties and decay rates compatible with the SM expectations~\cite{Langford:2021osp,ATLAS:2021vrm}, no fundamental principle excludes the existence of additional scalar bosons. 

In fact, resonant di-photon searches show intriguing indications for new Higgs bosons at the electroweak (EW) scale, both in the side-bands of SM Higgs analyses and in dedicated searches at 95$\,$GeV~\cite{LEPWorkingGroupforHiggsbosonsearches:2003ing,CMS:2018cyk,CMS:2022rbd,CMS:2022tgk,ATLAS:2023jzc} and 151.5$\,$GeV, with global significances of $3.8\sigma$ and $4.9\sigma$~\cite{Crivellin:2021ubm,Bhattacharya:2023lmu}, respectively. While the production mechanism for the 95$\,$GeV scalar candidate is currently unknown, data implies that the 151.5$\,$ GeV boson is mainly produced in association, in particular with missing energy~\cite{ATLAS:2021jbf}, but also leptons and ($b$-)jets~\cite{ATLAS:2023omk}. 

\begin{figure}[t!]
    \centering
        \includegraphics[width=0.49\linewidth]{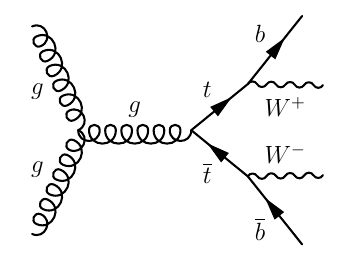}
    \includegraphics[width=0.49\linewidth]{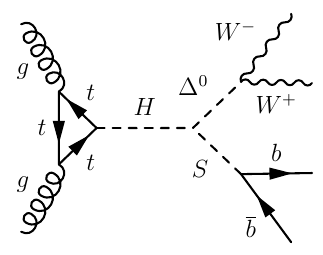}
    \caption{Feynman diagrams showing the leading SM contribution (left) to top pair production and decay as well as the NP effect mimicking this process (right).}
    \label{fig:NPttbar}
\end{figure}

The mechanism of associated production provides the connection to the ``LHC multi-lepton anomalies''~\cite{vonBuddenbrock:2016rmr,vonBuddenbrock:2017gvy,Buddenbrock:2019tua,vonBuddenbrock:2020ter,Hernandez:2019geu,Fischer:2021sqw}. These are processes involving multiple leptons and missing energy, with and without ($b$-)jets, with SM signatures such as $W^+ W^-$, $WWW$, $Wh$, $tW$, $t\bar t$, $t\bar tW$ and $t\bar t t\bar t$ (see Refs.~\cite{Fischer:2021sqw,Crivellin:2023zui} for a review) where deviations from the SM expectations have been observed. In particular, SM Higgs to $W^+ W^-$ analyses~\cite{CMS:2022uhn,ATLAS:2022ooq} prefer a new physics (NP) contribution, which is compatible with a $\approx$150\,GeV scalar that is produced in association with missing energy~\cite{vonBuddenbrock:2017gvy,Coloretti:2023wng}. Furthermore, the latest ATLAS analysis of the $t\bar t$ differential cross-sections~\cite{ATLAS:2023gsl} prefers a NP contribution from the decay of a heavy new scalar into two lighter ones, which subsequently decay into $W^+ W^-$ and $b\bar b$ (see Fig.~\ref{fig:NPttbar}), over the SM hypothesis by at least $5.8\sigma$~\cite{Banik:2023vxa}. Again, the masses are compatible with the hints for narrow resonances at 95\,GeV and 151.5\,GeV which decay dominantly to $b\bar b $ and $W^+ W^-$, respectively. Since there is no indication for a $151.5$\,GeV boson decaying to $ZZ$, this suggests that it is the neutral component of a real $SU(2)_L$ triplet. 

In this article, we propose a novel UV complete model that gives rise to these signatures and explains the $t\bar t$ differential distributions: the $\Delta$2HDMS. It contains two scalar doublets, a real singlet, and a real triplet Higgs. We will show that it predicts the experimentally preferred $\gamma\gamma$ signal strengths for the 95\,GeV and the 151.5\,GeV Higgs bosons and is compatible with the suggested signal in $W^+W^-$ (with a full jet veto) if one includes decays of the 95\,GeV state to invisible. Furthermore, it can account for the resonant $t\bar t$ excess at $\approx\! 400\,$GeV ($3.5 \sigma$ locally)~\cite{CMS:2019pzc} as well as the elevated four-top, $t\bar t W$ and $t\bar t Z$ cross sections and predicts a positive shift in the $W$ mass as suggested by the global fit.

\section{Model}

As outlined in the introduction, we propose the $\Delta$2HDMS model. It contains two Higgs doublets with hypercharge $Y=+1/2$ ($H_1$ and $H_2$) and a real singlet ($S$). This means it corresponds to a 2HDMS~\cite{He:2008qm,Grzadkowski:2009iz,Logan:2010nw,Boucenna:2011hy,He:2011gc,Bai:2012nv,He:2013suk,Cai:2013zga,Chen:2013jvg,Guo:2014bha,Wang:2014elb,Drozd:2014yla,Campbell:2015fra,Drozd:2015gda,vonBuddenbrock:2016rmr,Arhrib:2018qmw,Engeln:2020fld,Azevedo:2021ylf,Biekotter:2022abc}, plus  a real $SU(2)_L$ triplet ($\Delta$)~\cite{Ross:1975fq, Gunion:1989ci,Chankowski:2006hs,Blank:1997qa,Forshaw:2003kh, Chen:2006pb,Chivukula:2007koj,Bandyopadhyay:2020otm}\footnote{A model with the same field content has been studied in Ref.~\cite{Inoue:2015pza} where it was shown that it can lead to electroweak Baryogenesis. Supersymmetric versions of the model were considered in Refs.~\cite{Bandyopadhyay:2015tva,Bandyopadhyay:2015oga}. {Ref.~\cite{Inoue:2015pza} finds that electroweak Baryogenesis can be achieved in a wide range of parameters, including the one studied here. Note that our LHC analysis is fully complementary to the one of Ref.~\cite{Inoue:2015pza}: Ref.~\cite{Inoue:2015pza} proposed the two-step mechanism for electroweak Baryogenesis while we consider in detail the LHC phenomenology in light of the anomalies.}}. As in the 2HDMS, one can employ two $Z_2$ symmetries, one under which $S$ and $H_2$ are odd, and one under which $H_1$ and $\Delta$ are odd, while all other fields are even. This removes tri-linear coupling and thus most potential sources of CP-violation.\footnote{Alternatively, one can employ a $U(1)^\prime$ symmetry~\cite{Ko:2012hd,Banik:2023ecr}.} Furthermore, this allows only one term which contains an odd number of any field:\footnote{The terms $\mu H_1 H_2^\dagger$ softly breaks the $Z_2$ symmetries but is usually included for phenomenological reasons.}
\begin{equation}
    \mathcal{L}_{SH_1\Delta H_2 }=-\lambda S H_1^\dagger \Delta H_2+h.c.\,. \label{lambda}
\end{equation}

After electroweak symmetry breaking, in the approximation $v_\Delta,v_S,v_1\ll v_2$ (where $v_X$ is the vacuum expectation value of the field $X$), $H_1$, $H_2$, $S$ and $\Delta$ contain the mass eigenstates as given in Table~\ref{fields}. Note that the term in Eq.~(\ref{lambda}) does not lead to mixing among the scalars in this limit, but gives rise to the trilinear interactions $\mathcal{L}_{\phi^3}=-\frac{v}{4}\lambda \Delta^0 H S+\frac{v}{2}\lambda \Delta^+ H^- S+h.c.$ (assuming $\lambda$ to be real and $v=\sqrt{v_1^2+v_2^2}\approx v_2 \!\approx\! \sqrt{2} \, 174\,$GeV) where $\Delta$ decomposed according to the conventions of Ref.~\cite{Ashanujjaman:2023etj,Crivellin:2023xbu}. The former term leads to the desired decay of $H$ into $S\Delta^0$ needed to obtain the associated production mechanism. 

For the Yukawa Lagrangian, we take the right-handed up-type and down quarks to be odd under $Z_2^2$. Therefore, $H_2$ couples to all fermions, i.e.~the Yukawa sector is the one of a type-I 2HDM. However, we also assume that the $Z_2$ symmetries of the Yukawa Lagrangian are broken in the top sector by the term
\begin{equation}
    {-\mathcal{L}_Y^{\cancel{Z}_2}= \mu_t \sqrt{2}\dfrac{m_t}{v}  \bar Q_3 \widetilde H_1 u_3\,.}   
\end{equation} 
We defined this term such that $\mu_t$ is the relative coupling strength compared to the SM Higgs.\footnote{Note that if one defines this coupling in the down basis (i.e.~CKM rotations appear associated with left-handed up quarks), the charged Higgs does not generate flavour violation in the down-sector.} Therefore, $H$ and $A$ can be produced from this coupling via gluon fusion induced by a top-quark loop. Furthermore, this coupling will lead to the top-associated production of $H$ and $A$ (see Fig.~\ref{fig::CPodd}) as well as the $tb$ associated production of $H^\pm$. Finally, we add neutral fermions $\chi$ (singlets under the SM gauge group) with the Yukawa coupling $Y_\chi S\bar \chi\chi$ to get a non-zero decay rate of $S$ to invisible. 

\section{Phenomenology}

\begin{table}[!t]
\begin{center}
\begin{tblr}{colspec={Q[c,1.8cm]|Q[c,0.5cm]Q[c,1.2cm]Q[c,1.2cm]Q[c,1cm]}}
{}& $S$ &${{H_1}}$&${{H_2}}$&$\Delta$ \\
\hline
$SU( 2 )_L$&1&2&2&3\\
$U( 1 )_Y$& 0&{1/2}&{1/2}&0\\
$Z_2^1$& +&-&+&-\\
$Z_2^2$& -&+&-&+\\
components &$S$&$H,A,H^\pm$&$h,{G^0},{G^\pm }$&$\Delta^0,\Delta^ \pm$
\end{tblr}
\caption{Fields and their respective components in the limit of vanishing mixing angles. $h$ corresponds to the SM Higgs and $G^0$ and $G^\pm$ to the Goldstone bosons. Note that, following the common 2HDM conventions~\cite{Branco:2011iw}, the SM Higgs boson $h$ is contained within $H_2$ while the heavy doublet Higgs bosons $H$ and $A$ originate from $H_1$.}
\label{fields}
\end{center}
\end{table}

Here we study the phenomenology of our model in the limit of negligibly small mixing angles, i.e.~$v_S,v_\Delta,v_1\ll v_2$\footnote{Note that while $\Delta^0$ has a very small mixing with the SM Higgs boson due to the $Z_2$ symmetries, $S$ can in principle have a bigger mixing with $h$ and thus relevant gluon fusion and $Z$ strahlung production.}. Note that $v_\Delta$ is naturally small due to the additional symmetry of the potential in case it goes to zero. Furthermore, $v_\Delta$ leads to a positive-definite shift in the $W$ mass~\cite{Strumia:2022qkt,Dorsner:2007fy,FileviezPerez:2022lxp,Cheng:2022hbo,Rizzo:2022jti,Wang:2022dte,Chabab:2018ert,Shimizu:2023rvi,Crivellin:2023xbu}. In fact, $v_\Delta\approx2$\,GeV is preferred by the current global fit~\cite{Bagnaschi:2022whn,deBlas:2022hdk,ParticleDataGroup:2022pth} (driven by the CDF-II result~\cite{CDF:2022hxs}).\footnote{To be more specific, taking into account only the real Higgs triplet (i.e. disregarding the usually small effect of the mass splitting of the doublet components), the global fit~\cite{Bagnaschi:2022whn,deBlas:2022hdk,ParticleDataGroup:2022pth} including the CDF-II measurement~\cite{CDF:2022hxs} prefers $v_\Delta = 3.4 \pm 1.0$\,GeV, while excluding the CDF-II measurement one finds $v_\Delta = 2.3 \pm 1.7$\,GeV.}

In this limit, the only possible unsuppressed decays (i.e.~tree-level decays via order one couplings) of $H$ ($A$) are $t\bar t$ and $S\Delta^0$ ($t\bar t$, $ZH$ and $W^\pm H^\mp $) while $H^\pm$ decays dominantly to $t b$ and possibly to $\Delta^\pm S$.\footnote{For a dominant decay $H\to S\Delta^0$, a coupling $\lambda$ of order one is necessary. However, note that also the Higgs self-coupling in the SM is of order one and that this does not cause any theoretical problem related to perturbativity, unitarity, etc. Furthermore, the mass splitting among the components of the doublet is small enough to be consistent with perturbativity and vacuum stability as well~\cite{Broggio:2014mna}.} In the following, we will assume $m_S<m_{\Delta^0}\ll m_H<m_A \!\approx\! m_{H^\pm}$ with $m_H$ being below the top threshold. Note that a small value of  $v_\Delta$ is preferred by the average~\cite{deBlas:2022hdk,Athron:2022isz,Bagnaschi:2022whn} of the $W$ mass measurements~\cite{CDF:2022hxs,ATLAS:2023fsi,LHCb:2021bjt,CDF:2022hxs,ALEPH:2013dgf} (see Ref.~\cite{Chabab:2018ert,FileviezPerez:2022lxp, Cheng:2022hbo,Chen:2022ocr,Rizzo:2022jti,Chao:2022blc,Wang:2022dte,Shimizu:2023rvi,Lazarides:2022spe,Senjanovic:2022zwy,Crivellin:2023xbu,Chen:2023ins}). Therefore, $\Delta^0$ and $\Delta^\pm$ are quasi-degenerate and being triplet-like, decay dominantly to $W^+ W^-$ and $WZ$, respectively. Since $S$ is singlet-like, it decays mainly ($\!\approx\! 86\%$) to $b\bar b$ and has Br$(S\to \gamma\gamma)=1.4\times 10^{-3}$~\cite{LHCHiggsCrossSectionWorkingGroup:2016ypw,Braaten:1980yq, Sakai:1980fa,Inami:1980qp,Gorishnii:1983cu,Gorishnii:1990kd, Gorishnii:1990zu,Gorishnii:1991zr,Djouadi:1997rj,Degrassi:2005mc, Passarino:2007fp,Actis:2008ts,Djouadi:1990aj,Chetyrkin:1996sr,Baikov:2005rw,Spira:1991tj}. 

The main production mechanisms of $H$ and $A$ are gluon fusion and $t\bar t$ associated production, $H^\pm$ is produced in $tb$ associated production while $S$ and $\Delta^0$ ($\Delta^\pm$) are produced from the decay of $H$ ($H^\pm$). We fix $m_S$ and $m_{\Delta^0}$ (and therefore $m_{\Delta^\pm}$) to the values suggested by the indications for narrow resonances, i.e.~95\,GeV and 151.5\,GeV, respectively. 

\subsection{$t\bar t$ Differential Distributions}

Next, we consider the production of $H$ via top-quark-induced gluon fusion~\cite{LHCHiggsCrossSectionWorkingGroup:2016ypw,Graudenz:1992pv,Spira:1995rr,Anastasiou:2002yz,Harlander:2002wh,Harlander:2001is,Aglietti:2006tp,Li:2015kaa,Anastasiou:2006hc,Harlander:2005rq,Ravindran:2003um} and its subsequent decay to $\Delta^0S$. Note that this is the only unsuppressed decay of $H$ if its mass is below the $t\bar t$ threshold. With $S\to b\bar b$ and $\Delta^0\to W^+ W^-$, the resulting final state has the same signature as top-quark pair production and decay in the SM  (see Fig.~\ref{fig:NPttbar}). In fact, as shown in Ref.~\cite{Banik:2023vxa}, the tensions between the measured $t\overline{t}$ 
(normalized) differential cross-sections observed by ATLAS~\cite{ATLAS:2023gsl} and the corresponding SM predictions (which are compatible with the earlier CMS results~\cite{CMS:2018adi})\footnote{{These findings are also compatible with the latest differential $t\bar t$ results of CMS~\cite{CMS:2024ybg}. However, since we are performing a Monte-Carlo-based analysis, including the new CMS result would not have a significant impact because the SM prediction is the limiting factor and CMS performs a less detailed analysis than ATLAS in this respect. Furthermore, note that the $t\bar t$ differential distributions only provide the normalization of the total NP effect and a smaller effect would even be welcomed by data.}}, can be removed if this NP process is added to the SM \footnote{{Note that while in Ref.~\cite{Banik:2023vxa} we focus on the $m^{e\mu}$ and $|\Delta\phi^{e\mu}|$ distributions, we checked that the other (normalized) distributions which are in reasonable agreement with the SM predictions are only marginally affected by the NP contribution (see Fig.~4 of Ref.~\cite{Banik:2023vxa}).}}. Using the most conservative estimate for the NP effect, i.e.~the SM simulation among the set performed by ATLAS that is in the least tension with data\footnote{This is the Powheg+Pythia8 (rew) simulation, including the reweighting of the top $p_T$ by using the NLO predictions (see Ref.~\cite{ATLAS:2023gsl} for details).}, a $5.8\sigma$ preference for NP with a best fit value of \begin{equation}
    \sigma(pp\to H\to \Delta^0 S\to W^+ W^- b \bar b)\!\approx\! 5{\rm pb}\,,
\end{equation}
is found. Note that this result is approximately independent of $m_H$ for $250\,$GeV$<m_H<300$\,GeV. This translates to the preferred region (blue bands) for $\mu_t$ in Fig.~\ref{fig::Yt}.

\subsection{Resonant $t\bar t$, $t\bar t t\bar t$ and $t\overline{b}H^\pm\to t\overline{b}b\overline{t}$}

Because the CP-odd scalar $A$ is (mainly) contained in $H_1$ together with $H$, they both have the same coupling strength to top quarks. However, since $A$ couples axially, its production cross-section from gluon fusion is $\approx\! 1.5$ higher than the one of $H$ for equal masses (see, e.g., Refs.~\cite{Dawson:1998py,Spira:2016ztx}). If CP is conserved, as we assume, $A$ can neither decay to a pair of gauge bosons nor into two lighter Higgs bosons such that
\begin{align}
    &\Gamma(A\to t      \bar t) = \frac{3 \; G_F m_A  m_t^2 \mu_t^2}{4 \sqrt{2} \pi} \sqrt{1-4\dfrac{m_t^2}{m_A^2}} \nonumber\\
    &\Gamma(A\to HZ) = \frac{G_F m_A^3}{8 \sqrt{2} \pi} \lambda^{3/2}(m_H^2,m_Z^2;m_A^2) \nonumber\\
    &\Gamma(A\to H^{\pm}W^{\mp}) = \frac{G_F m_A^3}{4 \sqrt{2} \pi} \lambda^{3/2}(m_{H^\pm}^2,m_{W^\mp}^2;m_A^2)\\
    \nonumber & \text{with} \\
    & \lambda(x,y;z) = \left(1-\frac{x}{z}-\frac{y}{z}\right)^2 - 4\frac{xy}{z^2}
\end{align}
are the dominant decay modes, with $\tan(\beta) = v_2/v_1$ and $G_F$ being the Fermi constant.

\begin{figure}[t!] 
    \centering
    \includegraphics[width=0.49\linewidth]{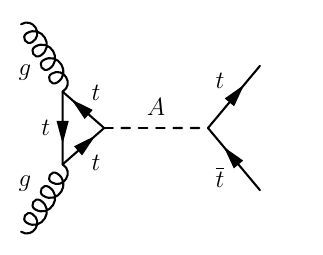}
    \includegraphics[width=0.49\linewidth]{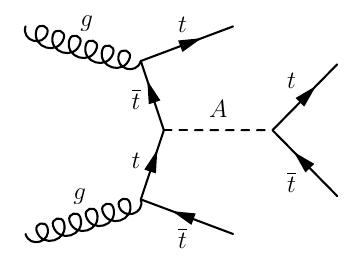}
    \caption{Feynman diagrams showing the production of $A$ via gluon fusion (left) and $t\bar t$ associated production (right) and its decays to top quarks.}
    \label{fig::CPodd}
\end{figure}

\begin{figure}[t!] 
    \centering
    \includegraphics[width=1\linewidth]{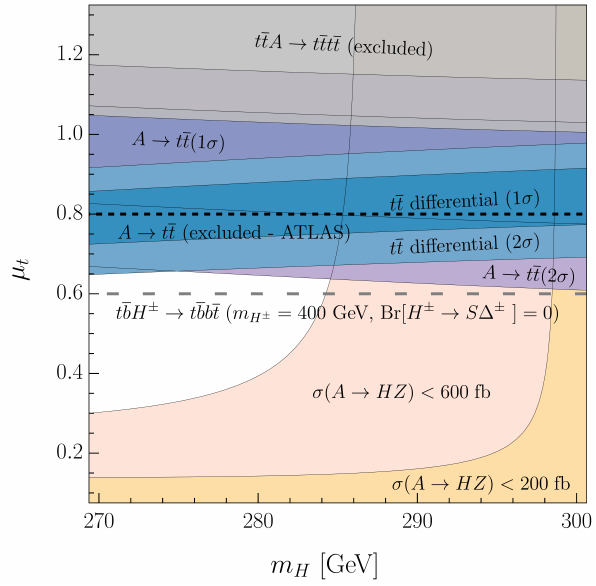}
    \caption{Constraints and preferred regions in the $m_H$-$\mu_t$ plane from $t\bar t$ differential distributions, resonant $t\bar t $  production and $t\bar t A\to t\bar tt\bar t$ (see main text for details).}
    \label{fig::Yt}
\end{figure}

To explain the CMS excess in resonant $t\overline{t}$ production at $\approx\!\! 400\,$GeV ($3.5 \sigma$ locally)~\cite{CMS:2019pzc} we set $m_A=400\,$GeV (see Fig~\ref{fig::CPodd}) and find\footnote{Note that CMS only reported the $p$-value as a function of $\mu_t$ while we assume a gaussian error for the cross-section (and thus $\mu_t^2$). The corresponding CMS analysis~\cite{ATLAS:2019npw} is done with less integrated luminosity and reaches only down to 500\,GeV. However, in the low mass region, a weaker-than-expected limit is observed.} 
\begin{equation}
 \mu_t^2 = (0.82\pm0.23)/{\rm Br}(A\to t\bar t)\,.
\end{equation}
This results in the violet band in Fig.~\ref{fig::Yt}.\footnote{Note that the differential lepton distributions of resonant $t\bar t$ production via $A$ are similar to the shapes within the SM.  Their effect drops out to a good approximation in the normalized differential distributions studied above.} However, ATLAS did not confirm this excess~\cite{ATLAS:2024vxm} putting an upper limit on $\mu_t$ of $\approx\!\! 0.8\,$at $95\%$ confidence level, as shown with a black dashed line in Fig.~\ref{fig::Yt}. 
The $t \overline{t}$ associated production of $A$ is constrained by the ATLAS limit on $\sigma(pp \to A t \overline{t} \to t\bar t t\bar t$)~\cite{ATLAS:2022rws} (see Fig.~\ref{fig::CPodd}). Here, a 2HDM of type II resulted in a $95 \%\;$CL lower limit of $1 \lesssim \tan\beta $ for $m_A \!\approx\! 400$\,GeV and thus $\mu_t\lesssim 1/{\rm Br}(A\to t\bar t)$ (gray excluded region in Fig.~\ref{fig::Yt}). However, also the total four-top cross-section $\sigma(pp\to t\bar tt\bar t)$ at 13\,TeV is measured by ATLAS~\cite{ATLAS:2023ajo} and CMS~\cite{CMS:2023ftu} to be $22.5^{+6.6}_{-5.5}\;$fb and $17.7^{+6.0}_{-5.4}\;$fb, respectively. In our model, the NP contribution to the total $t\bar tt\bar t$ cross section induced by $pp \to A t \bar t \to t\bar t t\bar t$ can be obtained from the ATLAS analysis of $\sigma(pp \to A t \bar t \to t\bar t t \bar t)$: for $m_A \!\approx\! 400$\,GeV, the $95 \%\;$CL upper limit on the total cross-section of $14.4\;$fb, corresponds to $\tan(\beta) > 1$ (see Fig.~8 in Ref.~\cite{ATLAS:2022rws}). Therefore, $\sigma(pp \to A t \bar t \to t\bar tt\bar t) \!\approx\! 14\mu_t^2\;$fb and thus nicely compatible with the experimentally preferred NP contribution to the cross-section.

ATLAS searches for the process $pp\to A\to ZH$ with $H\to t\bar t$~\cite{ATLAS:2023zkt}. At the point in parameter space closest to our mass region, i.e.~$m_A \!\approx\! 450\,$GeV and $m_H \!\approx\! 350\,$GeV, a limit of $\!\approx\! 600\;$fb is observed, compared to an expected limit of $\!\approx\! 400\;$fb. Even though we have $H\to W^+ W^- b \bar b$ instead of $H\to t\bar t$, the signatures are similar and one can use these results to estimate a limit on $pp\to A\to ZH$ in our model (see yellow and pink region in Fig.~\ref{fig::Yt}).\footnote{Depending on $m_H$, we can have the decay $A\to ZH$ with $H\to \Delta^0 S \to  W^+ W^- b\bar b$ which could potentially contaminate the $t\bar t Z$ cross-section measurement. The most precise measurement of the inclusive $t\bar t Z$ cross-section to date stems from the ATLAS experiment~\cite{CERN-EP-2023-252}, which is consistent with the SM. That said, all channels, except for one (four-leptons with different flavor) rely on the use of Machine Learning. This will suppress the sensitivity to non-$t\bar t Z$ final states, such as $A\to ZH, H\rightarrow W^+W^-b\bar b$, in particular, for $m_H < m_{t\bar t}$.} Note that the top associated production of $H$ leads to an effect in $t\bar tW$. While this effect goes in the direction preferred when comparing the SM prediction to the experimental value, it only gives a moderate effect. 

CMS~\cite{CMS:2019rlz} and ATLAS~\cite{ATLAS:2021upq} searched for $t\overline{b}H^\pm\to t\overline{b}b\overline{t}$. They found a lower bound on $\tan(\beta)$ (corresponding to an upper bound on $Y_t^2 = 1/\tan^2(\beta)$) as a function of $m_{H^\pm}$. For $m_{H^\pm}$ the upper limit on $\mu_t$ is around $0.6$. However, depending on the coupling $\lambda$, a sizable decay rate for $H^\pm\to \Delta^\pm S$ is possible. With $\Delta^\pm\to W^\pm Z$ and $S\to b\bar b$, this results in $t\bar t Z$ and $t\bar t W$-like siganture. In fact, given that there is a preference for a $\approx\!\! 200\;$fb contribution in the latest $t\bar tW$ analysis of ATLAS~\cite{ATLAS:2023gon}, a NP contribution of around $\approx\!\! 400\;$fb is allowed at the $2\sigma$ level. Since the production cross section for $t b H^\pm\to t \bar b b \bar t$ is $600\;$fb for $m_{H^\pm}=400\,$GeV and $\mu_t=1$, it is possible to evade the $t\overline{b}H^\pm\to t\overline{b}b\overline{t}$ bound, which is therefore only shown as a gray dashed line in Fig.~\ref{fig::Yt}.

\subsection{$151.5\,$GeV and $95\,$GeV signal strength}

\begin{figure}[t]
    \includegraphics[width=1\linewidth]{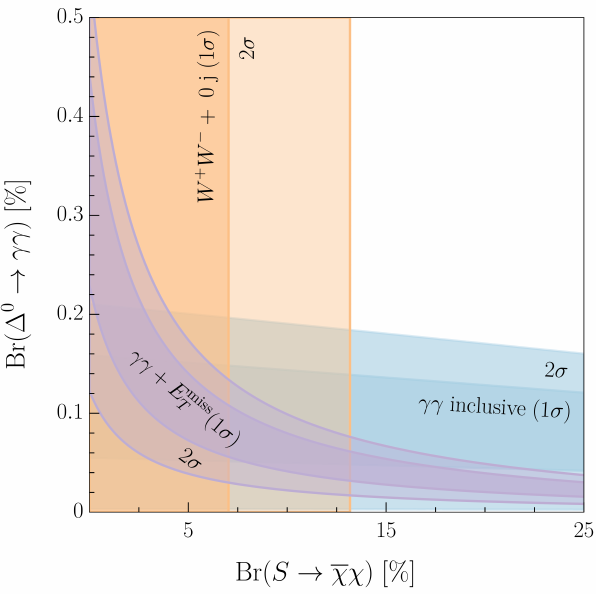}
    \caption{Preferred regions from signal strength measurements of $\gamma\gamma$ (inclusive), $\gamma\gamma+(E_T^{\text{miss}}>90\,$GeV$)$ and $W^+ W^-$ (jet veto) of the 151.5\,GeV boson $\Delta^0$ as a function of its branching ratio to $\gamma \gamma$ and the branching ratio of the $95\,$GeV boson $S$ to invisible, for $m_H=290\,$GeV.}
    \label{S152SigmalStrength}
\end{figure}

The main indications for a new resonance at $\approx$151.5\,GeV are in inclusive $\gamma\gamma$ searches~\cite{CMS:2021kom,ATLAS:2021jbf}, $\gamma\gamma+(E_T^{\text{miss}}>90$\,GeV)~\cite{ATLAS:2020pvn,CMS:2018nlv} and $W^+ W^-$ (without jets)~\cite{CMS:2022uhn,ATLAS:2022ooq}, where $E_T^{\text{miss}}$ stands for missing transverse energy. Ref.~\cite{Crivellin:2021ubm} found $\sigma(pp\to \Delta^0\to \gamma\gamma)\!\approx\! (6\pm3)$\,fb and $\sigma(pp\to \Delta^0\to \gamma\gamma+(E_T^{\text{miss}}>\!90{\rm GeV}))\!\approx\!(0.6\pm 0.2)$\,fb. The latter is obtained within the simplified model $pp\to (270\,{\rm GeV})\to (151.5\,{\rm GeV}\to \gamma\gamma)(151.5^*\,{\rm GeV}\to \bar \chi\chi)$ where $m_{\chi}=50\,$GeV. The process $W^+ W^-$ with a full jet veto was studied for the direct production of a new scalar in Ref.~\cite{Coloretti:2023wng} finding $\approx\!\! 2\sigma$ significance for a non-zero cross-section of $\approx\!1$pb at $\approx\!\!151.5\,$GeV. In our model, $\Delta^0$ is produced in association with $S$ and therefore mainly with $b$-jets, taus and leptons. Thus, a non-zero branching ratio of $S\to\;$invisible, Br$(S\to  \bar \chi  \chi)$, is needed to pass the full jet veto and to give rise to $\gamma\gamma+E_T^{\text{miss}}$. We simulated both $\gamma\gamma+E_T^{\text{miss}}$ and $W^+ W^-+E_T^{\text{miss}}$ in our model to obtain the efficiency and the significances for $S\to E_T^{\text{miss}}$, i.e.~$m_\chi<95/2\,$GeV.\footnote{For the simulations, we employed {\tt MadGraph5aMC@NLO}~\cite{Alwall:2014hca} with the parton shower performed by {\tt Pythia8.3}~\cite{Sjostrand:2014zea} and the detector simulation carried out with {\tt Delphes}~\cite{deFavereau:2013fsa}. We validated our analysis of $\gamma\gamma+E_T^\text{miss}$ by comparing it against the ATLAS simulation in Ref.~\cite{ATLAS:2021jbf}. To correct for our fast simulation vs the full detector simulation of ATLAS, we added a 15\,GeV smearing on the missing transverse energy (excluding jets) and on the jet energies. Note that also Drell-Yan production, i.e.~$pp\to W^*\to \Delta^\pm \Delta^0$ with $\Delta^0\to ZW$ and $Z\to \nu\nu$ and/or $W\to\ell\nu$, contributes to $\gamma\gamma+(E_T^\text{miss}>90)$ which is responsible for a non-vanising cross-section at Br$(S\to \bar \chi\chi)=0$ in Fig.~\ref{S152SigmalStrength}. } The results for the three regions $\gamma\gamma$ (inclusive), $\gamma\gamma+(E_T^{\text{miss}}>90\,$GeV$)$ and $W^+ W^-+0$~jets are shown in Fig.~\ref{S152SigmalStrength} for $m_H=290\,$GeV. One can see that all three signal strengths are nicely compatible for Br$(S\to  \bar \chi \chi)
\!\approx\!5\%$ and Br$(\Delta^0\to  \gamma\gamma)\!\approx\!0.1\%$. Note that the latter value is within the natural expectations for the scalar triplet with hypercharge~$Y=0$~\cite{prepar:preparation}.

Finally, the di-photon signal strength for $S$ ($\approx95$\,GeV) induced via $pp\to H\to \Delta^0 S$~\cite{Banik:2023vxa} of $\mu_{\gamma\gamma}\!\approx\!0.1$ (w.r.t.~a hypothetical SM Higgs-like boson with the same mass) is at the low side of the preferred range $\mu_{\gamma\gamma}=0.24^{+0.09}_{-0.08}$~\cite{Biekotter:2023oen}. However, one could allow for a moderate mixing angle with the SM Higgs boson of around $0.2$, which is possible if $v_S$ is non-negligible. In this setup, one would be close to the central value while being consistent with SM Higgs boson signal strength measurements and at the same time explaining (partially) the LEP excess in $Z$ strahlung as well as the indications for $\tau\tau$~\cite{CMS:2022goy} and $W^+W^-$~\cite{Coloretti:2023wng}.

\section{Conclusions and Outlook}
\label{sec:conclusions}

We propose the $\Delta$2HDMS as a combined explanation of the LHC multi-lepton anomalies, the resonant di-photon ($\approx$95\,GeV and $\approx$151.5\,GeV) and top quark excesses ($\approx$400\,GeV) with significances of $>\!5\sigma$, $3\sigma\!-\!5\sigma$ and $3.4\sigma$, respectively. The model is obtained by extending the SM with a second Higgs boson doublet as well as a real $SU(2)_L$ singlet and a real $SU(2)_L$ triplet and naturally predicts (at tree-level) a positive shift in the $W$ mass in agreement with the current global fit for $v_\Delta=O({\rm 1 GeV})$. In particular, a benchmark points with
\begin{equation}
\begin{aligned}
    m_{\Delta^\pm}&\approx m_{\Delta^0}=151.5\,{\rm GeV},\; m_{S}=95\,{\rm GeV},\\m_{H^\pm}&\approx m_{A}\approx 400\,{\rm GeV}, v_\Delta \approx 2\,{\rm GeV},\; v_S=O({\rm GeV}),\\
    m_H&=290\,{\rm GeV},\; \mu_t=0.7,\; \lambda=O(1),\; \tan\beta\gtrapprox5,
\end{aligned}
\end{equation}
is consistent with all signals and bounds.

While we provided proof of principle that all these anomalies can be explained without violating other bounds, in the future, a detailed examination of the NP effects in the differential $t\bar t W$ and $t\bar t Z$ distributions would be very interesting. Additionally, note that e.g.~the disappearance of the resonant $t\bar t$ would not falsify the model: if $A$ were lighter, i.e.~below the top threshold, it could contribute to the differential $t\bar t$ distributions via $pp\to A\to W^\pm H^\mp$ with $H^\pm\to tb$, which would supplement the effect of $pp\to H\to \Delta^0S\to W^+ W^- b \bar b$. This would allow for an explanation of the anomaly with a small top Yukawa coupling and $H$ could be lighter since Br$(A\to ZH)$ would be smaller, and thus the $t\bar t Z$-like signal is reduced.

\begin{figure}[t]
    \includegraphics[width=1\linewidth]{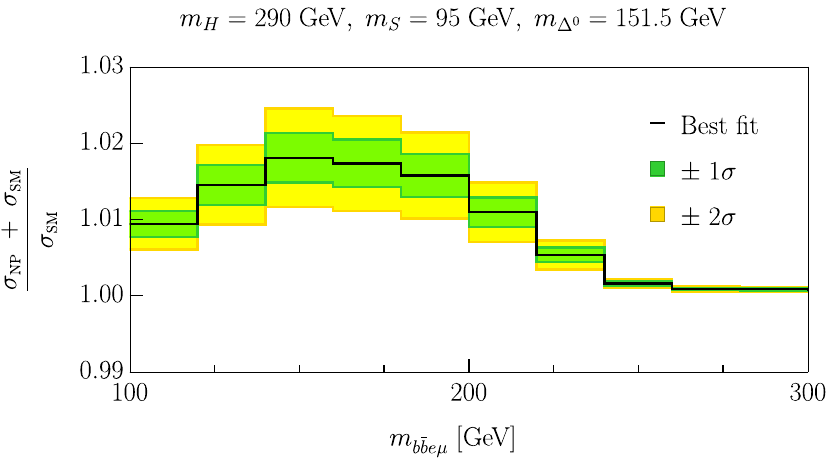}
    \caption{Predicted deviations from the SM predictions in the invariant mass distribution of the $b\bar b e\mu $ system for our benchmark point. Note that these results are in agreement with the preliminary ATLAS results of Ref.~\cite{Rottler:2023xwl}. }
    \label{mbbll}
\end{figure}

Our model can be tested in multiple ways with forthcoming measurements. For example, it predicts deviations from the SM predictions in the invariant mass of the $b\bar b e\mu$ system in the differential $t\bar t$ analyses (see Fig.~\ref{mbbll}) and in asymmetric Higgs signals, e.g. in $\gamma\gamma bb$. Furthermore, it opens a window of opportunity for the exploration and observation of new bosons at the EW scale at future $e^+e^-$ accelerators, in particular the charged component of the triplet ($\Delta^\pm$), such as the Circular Electron-Positron Collider (CEPC)~\cite{CEPCStudyGroup:2018ghi,An:2018dwb}, the Compact Linear Collider (CLIC)~\cite{CLICdp:2018cto}, the Future Circular Collider (FCC-ee)~\cite{FCC:2018evy,FCC:2018byv} and the International Linear Collider (ILC)~\cite{ILC:2013jhg,Adolphsen:2013jya}.

\begin{acknowledgments}
We thank Saiyad Ashanujjaman, Sumit Banik and Gino Isidori for useful discussions. The work of A.C.~and G.C.~are supported by a professorship grant from the Swiss National Science Foundation (No.\ PP00P2\_211002). B.M.~gratefully acknowledges the South African Department of Science and Innovation through the SA-CERN program, the National Research Foundation, and the Research Office of the University of the Witwatersrand for various forms of support.
\end{acknowledgments}

\bibliographystyle{utphys}
\bibliography{PRD_resubmission}

\end{document}